# Electric field modulation of thermopower for transparent amorphous oxide thin film transistors

Hirotaka Koide[1], Yuki Nagao[1], Kunihito Koumoto[1], Yuka Takasaki[1], Tomonari Umemura[1], Takeharu Kato[2], Yuichi Ikuhara[2,3], and Hiromichi Ohta[1,4,a]

[1]*Graduate School of Engineering, Nagoya University, Chikusa, Nagoya 464–8603, Japan*

[2]*Nanostructures Research Laboratory, Japan Fine Ceramics Center, 2–4–1 Mutsuno, Atsuta, Nagoya 456–8587, Japan*

[3]*Institute of Engineering Innovation, The University of Tokyo, Bunkyo, Tokyo 113–8656, Japan*

[4]*PRESTO, Japan Science and Technology Agency, Sanbancho, Tokyo 102–0075, Japan*

[a]Correspondence should be addressed H.O. (h-ohta@apchem.nagoya-u.ac.jp)

**To clarify the electronic density of states (DOS) around the conduction band bottom for state of the art transparent amorphous oxide semiconductors (TAOSs), InGaZnO$_4$ and In$_2$MgO$_4$, we fabricated TAOS-based transparent thin film transistors (TTFTs) and measured their gate voltage dependence of thermopower (*S*). TAOS-based TTFTs exhibit an unusual *S* behavior. The |*S*|-value abruptly increases, but then gradually decreases as $V_g$ increases, clearly suggesting the anti-parabolic shaped DOS is hybridized with the original parabolic shaped DOS around the conduction band bottom.**

Transparent amorphous oxide semiconductors (TAOSs) based transparent thin film transistors (TTFTs) have attracted increased attention as active switching devices for flat panel displays such as AMLCDs and AMOLEDs. TAOS-TFFTs can be fabricated at low temperature, do not show a visible light response, and exhibit a rather large carrier mobility ($\mu$ ~10 cm$^2$V$^{-1}$s$^{-1}$) compared to hydrogenated amorphous Si-TFTs ($\mu$ ~1 cm$^2$V$^{-1}$s$^{-1}$).[1] To date, a number of TAOSs, including InGaZnO,[1–3] InZnO,[4] SnGaZnO,[5] ZrInZnO,[6] AlZnSnO,[7] ZnInSnO,[8] HfInZnO,[9] MgInZnO,[10] AlInSnO,[11] and AlSnZnInO[12] systems have been proposed for the active channel layer of TTFTs.

Our preliminary study revealed that In$_2$MgO$_4$, which is a wide bandgap (~3.4 eV) oxide semiconductor[13, 14] in the crystalline phase, exhibits excellent transistor characteristics. Mg is an abundant element; the weight percentage of Mg on the Earth's surface is 1.93%, which is three orders of magnitude larger than that of Ga (0.001%) or Zn (0.004%). In this letter, we examine In$_2$MgO$_4$ as well as InGaZnO$_4$ as TAOSs for TTFTs.

In TAOS-TTFTs, transistor characteristics are strongly affected by the presence of extra density of states (DOS) such as tail and/or subgap states,[15–17] which are caused by structural randomness, defects, impurities, etc. These extra DOS, which have been modeled by device simulation[16] and hard X-ray photoelectron spectroscopy,[17] should play the essential role of transistor switching of TAOS-based TTFTs.

To further clarify the electronic DOS around the conduction band bottom for state of the art TAOSs, $InGaZnO_4$ and $In_2MgO_4$, we fabricated TAOS-based transparent thin film transistors (TTFTs) and measured their gate voltage dependence of thermopower ($S$). If a semiconductor has a simple parabolic shaped conduction band DOS, then the absolute value of thermopower $|S|$ monotonically decreases with as $V_g$ increases[18,19] because the $S$-value reflects the energy differential of the DOS at the Fermi energy: $\left[\frac{\partial DOS(E)}{\partial E}\right]_{E=E_F}$. Thus, the gate voltage dependence of $|S|$ for transistors is a good measure to clarify the electronic DOS for semiconductors. Herein we report the fabrication process and gate voltage dependences of $S$ for two TAOS-based TTFTs, $In_2MgO_4$ and $InGaZnO_4$. TAOS-based TTFTs exhibit an unusual $S$-behavior, which suggests that the anti-parabolic shaped extra state plays an essential role in transistor switching of the TAOS-based TTFTs.

We fabricated TAOS-based TTFTs on an alkaline-free glass substrate (0.5 mm$^t$, Corning EAGLE 2000) using the following procedure. First, a 100 nm thick ITO (10 wt% $SnO_2$ doped $In_2O_3$) film, which served as the bottom gate electrode, was deposited by pulsed laser deposition (PLD, KrF excimer laser, 10 Hz, ~1 Jcm$^{-2}$pulse$^{-1}$, oxygen pressure 3 Pa) onto the glass substrate. Next, a 300 nm thick Y$_2$O$_3$ gate dielectric film was deposited by PLD (~2 Jcm$^{-2}$pulse$^{-1}$, oxygen pressure 1 Pa) through a shadow mask. The dielectric permittivity ($\varepsilon_r$) of the Y$_2$O$_3$ film was 20, which was measured by an LCR meter (4284A, Agilent). Then, a 40 nm thick In$_2$MgO$_4$ or InGaZnO$_4$ film (surface area: 800 μm × 1000 μm) was deposited through a stencil mask by PLD (~3 Jcm$^{-2}$pulse$^{-1}$, oxygen pressure 1 Pa) onto the Y$_2$O$_3$ / ITO bilayer laminate.

The metal composition ratio, which was measured using the double focusing magnetic sector field ICP–MS (ELEMENT 2, Thermo Scientific), of In:Mg was 2.2:1 for the In$_2$MgO$_4$ film and In:Ga:Zn was 1:1.07:0.79 for the InGaZnO$_4$ film, indicating that the film composition was slightly In-rich. Finally, 20 nm thick ITO films (surface area: 400 μm × 400 μm), which were used as the source and drain electrodes, were deposited by PLD as described above. The deposition rates for the ITO, Y$_2$O$_3$, In$_2$MgO$_4$, and InGaZnO$_4$ films were ~5, ~5, ~10 and ~10 nm·min$^{-1}$, respectively. All the deposition processes were performed at room temperature (substrate was not heated). After the deposition processes, the devices were annealed at 400 °C (In$_2$MgO$_4$) or 300 °C (InGaZnO$_4$) for 30 min in air.

Figure 1 shows a TAOS-based TTFT. The resultant TAOS-based TTFTs were fully transparent in the visible light region, as shown in Fig. 1(a). The channel length ($L$) and channel width ($W$) were both 400 μm [Fig. 1(b)]. Additionally, the featureless structure of the In$_2$MgO$_4$ was observed in a cross sectional transmission electron microscope image of the resultant TAOS (In$_2$MgO$_4$)-based TTFT (HRTEM, TOPCON

EM-002B, acceleration voltage 200 kV, TOPCON) [Fig. 1(c)].

The transistor characteristics of the TAOS-TTFTs were measured by a semiconductor device analyzer (B1500A, Agilent) at room temperature in a sealed box. The *S*-values were measured by the conventional steady state method as follows.[18,19] Two K-type thermocouples were in contact with both channel edges. Two Peltier devices placed under the device were used to introduce a temperature difference ($\Delta T_{max}$ = 5 K) between the source and drain electrodes. Then the thermo-electromotive force ($V_{TEMF}$) was measured during the $V_g$-application. The values of *S* were obtained from the slope of $V_{TEMF}$–$\Delta T$ plots (data not shown).

Figure 2 shows (a) typical transfer ($I_d$–$V_g$) and (b) the output ($I_d$–$V_d$) characteristics of the TAOS-TTFTs. The drain current ($I_d$) of the TTFTs increased markedly as the gate voltage ($V_g$) increased; hence, the channels in both TTFTs were *n*-type, and electron carriers accumulated by positive $V_g$ application. The $I_d^{0.5}$ vs. $V_g$ plots indicated that $V_{gth}$ for In$_2$MgO$_4$ was −0.74 V while for InGaZnO$_4$ was 0.06 V. Table I summarizes the on−off current ratio ($V_{gth}$), subthreshold swing factor (*S.S.*), and saturation mobility ($\mu_{sat}$) for the TAOS-TTFTs. The $\mu_{sat}$ values were obtained by analyzing the $I_d$−$V_g$ curves in the saturation region using the following equation

$$I_{dsat} = \frac{WC_i}{2L} \mu_{sat} \left(V_g - V_{gth}\right)^2$$

where $C_i$ and $V_{gth}$ denote capacitance per unit area (58 nFcm$^{-2}$) and threshold gate voltage, respectively. Both In$_2$MgO$_4$- and InGaZnO$_4$-based TTFTs exhibited a clear pinch-off and current saturation in $I_d$ as shown in Fig. 3 (b), indicating that the operation of this TTFT conformed to standard FET theory. These results clearly indicate that In$_2$MgO$_4$- and InGaZnO$_4$-based TTFTs have similar transistor characteristics.

We then measured the gate voltage dependence of *S* for the TAOS-TTFTs to clarify the electronic DOS around the conduction band bottom. Figure 3 shows the gate voltage dependences of |*S*| for the In$_2$MgO$_4$- and InGaZnO$_4$-based TTFTs at room temperature. The *S* values were negative, confirming that the channel was an *n*-type. For both In$_2$MgO$_4$- and InGaZnO$_4$-based TTFTs, the |*S*|-value abruptly increased and then gradually decreased as $V_g$ increased. The slope of the |*S*|–log ($V_g$–$V_{gth}$) plots (shown in the inset) in the higher $V_g$–$V_{gth}$ region (>3 V) corresponded −198 μVK$^{-1}$, indicating that the original conduction band DOS of the TAOSs is parabolic shaped. On the other hand, the positive slope of the |*S*|–log ($V_g$–$V_{gth}$) plots in the lower $V_g$–$V_{gth}$ region (<3 V) indicated an anti-parabolic shaped DOS as schematically illustrated in Fig. 4. Additionally, the present results supported that the anti-parabolic shaped extra state was hybridized just below the original conduction band bottom of the TAOS-based TTFT channel, which is similar device simulation model[16] and hard X-ray photoelectron spectroscopy.[17] Thus, the anti-parabolic DOS plays an essential role in transistor switching of TAOS-based TTFTs.

In summary, to clarify the electronic DOS around the conduction band bottom for TAOSs, we have demonstrated fabrication and thermopower (*S*) modulation of TAOS-based TTFTs using two state of the art TAOSs, In$_2$MgO$_4$ and InGaZnO$_4$. TAOS-based TTFTs exhibit an unusual *S* behavior. The |*S*|-value abruptly increases and then gradually decreases as $V_g$ increases, clearly suggesting the anti-parabolic shaped DOS is hybridized with the original parabolic shaped DOS around the conduction band bottom. These results support that the anti-parabolic shaped extra state plays an essential role in transistor switching of TAOS-based TTFTs.

The authors would like to thank K. Nomura for valuable discussions and A. Okada for supplying alkaline free glass substrates. A part of this work was financially supported by a Grant-in-Aid for Scientific Research (No. 22015009, 22360271) from the Ministry of Education, Culture, Sports, Science, and Technology of Japan.

Table I  On–off current ratio, threshold gate voltage ($V_{gth}$), subthreshold swing factor (*S.S.*), and saturation mobility ($\mu_{sat.}$) for TAOS-based TFTs.

|  | ON/OFF | $V_{gth}$ (V) | *S.S.* (mV·decade$^{-1}$) | $\mu_{sat.}$ (cm$^2$V$^{-1}$s$^{-1}$) |
|---|---|---|---|---|
| In$_2$MgO$_4$ | >10$^7$ | –0.75 | 150 | 4.6 |
| InGaZnO$_4$ | >10$^7$ | 0.065 | 500 | 6.3 |

H. Koide *et al.***FIG. 1 (Color online)** TAOS-based transparent thin film transistor (TTFT). (a) Photograph of TAOS-TTFTs. Twenty TTFTs were fabricated on an alkaline-free glass substrate. (b) Optical micrograph of the TAOS-TTFTs with channel lengths and widths of 400 μm. Two K-type thermocouples are visible. (c) Cross-sectional TEM image of an $In_2MgO_4$-based TTFT. Featureless structure of $In_2MgO_4$ is shown.

**FIG. 2 (Color online)** Transistor characteristics of the TAOS-TTFTs at room temperature. (a) Transfer ($I_d$–$V_g$) characteristics at $V_d$ = 5 V. (b) Output ($I_d$–$V_d$) characteristics (red: $InGaZnO_4$, blue: $In_2MgO_4$). Clear pinch-off and current saturation are seen.

**FIG. 3 (Color online)** Gate voltage dependence of thermopower for the TAOS-TTFTs. |$S$|-value abruptly increases and then gradually decreases as $V_g$ increases. Inset shows |$S$| vs. log ($V_g$–$V_{gth}$) plots. Slope of the |$S$|–log ($V_g$–$V_{gth}$) plots in the higher $V_g$–$V_{gth}$ region (>3 V) corresponds −198 μVK$^{-1}$, indicating that the original conduction band DOS of the TAOSs is parabolic. On the other hand, the positive slope of |$S$|–log ($V_g$–$V_{gth}$) plots in the lower $V_g$–$V_{gth}$ region (<3 V) indicates an anti-parabolic shaped DOS.

**FIG. 4 (Color online)** Schematic DOS–$V_g$ model for the TAOS-TTFTs. Anti-parabolic shaped extra DOS is hybridized just below the original DOS of the TAOS-based TTFT channel.

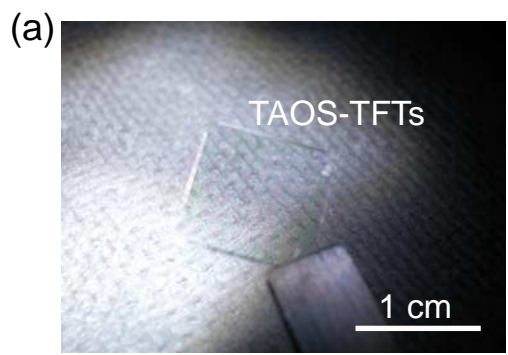
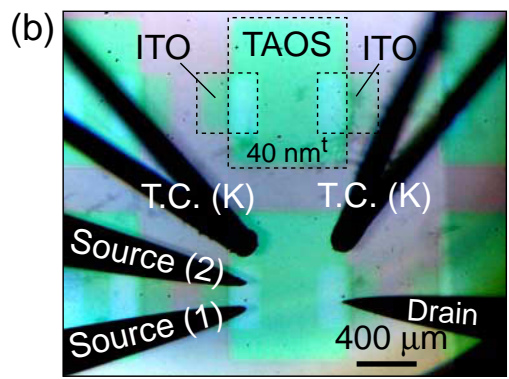
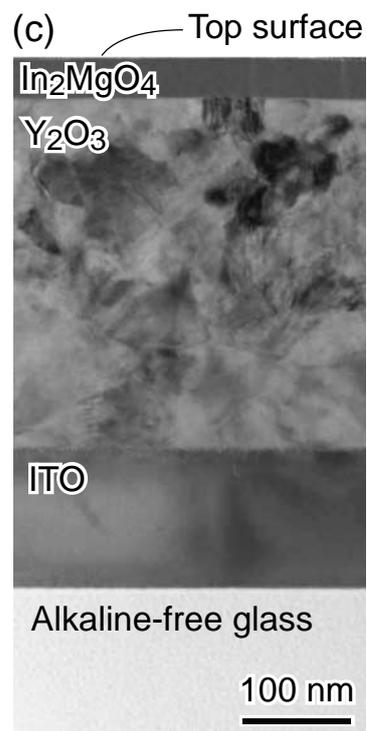

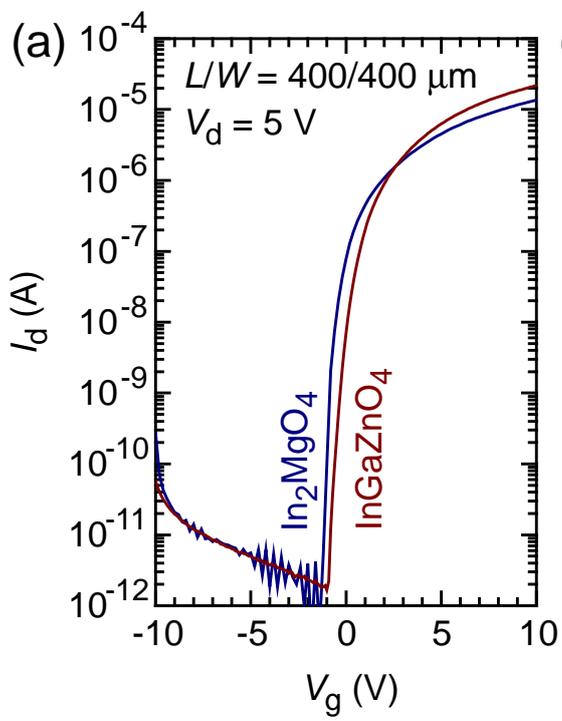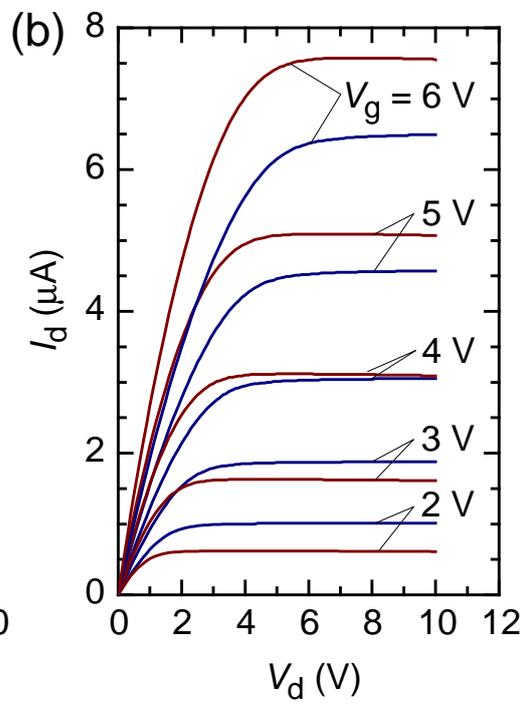

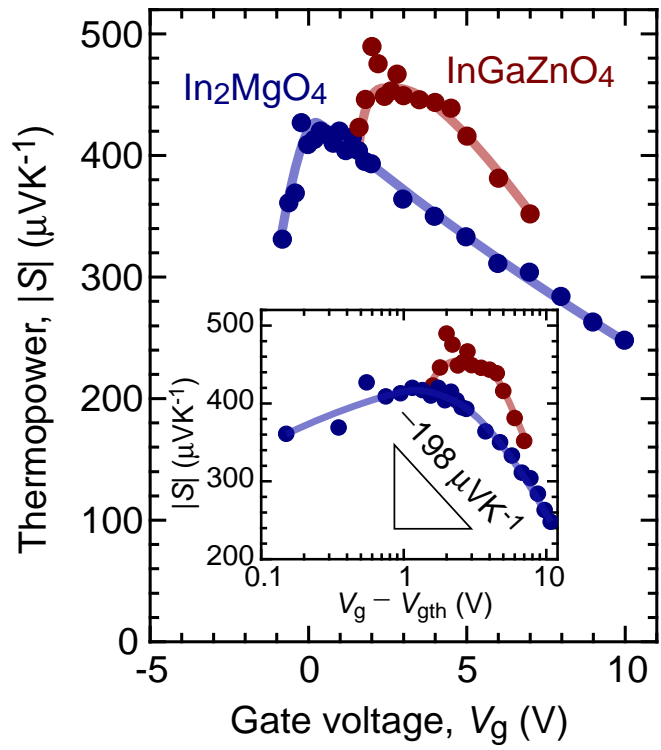

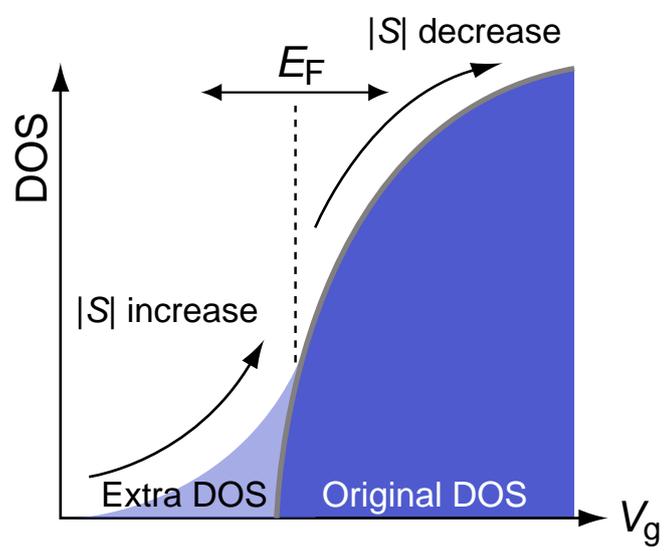